\documentclass{PoS}

\usepackage{amsmath}
\usepackage{slashed}
\usepackage{newtxtext,newtxmath}

\newcommand{\eff}{{\textrm {eff}}}

\newcommand{\rad}{{\textrm {rad}}}

\newcommand{\GeV}{{\textrm {GeV}}}
\newcommand{\EW}{{\textrm {EW}}}
\newcommand{\SNR}{{\textrm {SNR}}}
\newcommand{\ggF}{{\textrm {ggF}}}
\newcommand{\VBF}{{\textrm {VBF}}}
\newcommand{\SM}{{\textrm {SM}}}
\newcommand{\mmin}{{\textrm {min}}}
\newcommand{\BH}{{\textrm {BH}}}
\newcommand{\Planck}{{\textrm {Planck}}}

\title{Gravitational waves and collider signatures from holographic phase transitions in soft walls}

\ShortTitle{Gravitational waves and collider signatures from holographic phase transitions}

\author{\speaker{Eugenio Meg\'{\i}as}$\;^{a,b}\,$, Germano Nardini$\;^{c}$ and Mariano Quir\'os$\;^{d,e}$~\thanks{The work of EM is supported by the Spanish MINEICO and European FEDER funds (grants FPA2015-64041-C2-1-P and FIS2017-85053-C2-1-P), Junta de Andaluc\'{\i}a (grant FQM-225), and Basque Government (grant IT979-16). The research of EM is also supported by the Ram\'on y Cajal Program of the Spanish MINEICO (grant RYC-2016-20678), and by the Universidad del Pa\'{\i}s Vasco UPV/EHU, Bilbao, Spain, as a Visiting Professor. The work of MQ is partly supported by Spanish MINEICO (grants CICYT-FEDER-FPA2014-55613-P and FPA2017-88915-P), by the Catalan Government under grant 2017SGR1069, and Severo Ochoa Excellence Program of MINEICO (grant SEV-2016-0588).} 
  \\
        \llap{$^a$} Departamento de F\'{\i}sica At\'omica, Molecular y Nuclear and \\ Instituto Carlos I de F\'{\i}sica Te\'orica y Computacional, Universidad de Granada, \\ Avenida de Fuente Nueva s/n,  18071 Granada, Spain \\
        \llap{$^b$} Departamento de F\'{\i}sica Te\'orica, Universidad del Pa\'{\i}s Vasco UPV/EHU, \\ Apartado 644, 48080 Bilbao, Spain \\
        \llap{$^c$} Faculty of Science and Technology, University of Stavanger, 4036 Stavanger, Norway\\
        \llap{$^d$} Department of Physics, University of Notre Dame, 225 Nieuwland Hall,\\ Notre Dame, IN 46556, USA\\
                \llap{$^e$} Institut de F\'{\i}sica d'Altes Energies (IFAE), The Barcelona Institute of  Science and Technology (BIST), Campus UAB, 08193 Bellaterra (Barcelona) Spain\\
      E-mail: \email{emegias@ugr.es, germano.nardini@uis.no, quiros@ifae.es}}

\abstract{Using a 
five-dimensional 
warped model including a scalar potential with an exponential behavior in the infrared, and strong back-reaction over the metric, we study the electroweak phase transition, and explore parameter regions that were previously inaccessible. The model exhibits gravitational waves and predicts a stochastic gravitational wave background observable, both at the Laser Interferometer Space Antenna and at the Einstein Telescope. Moreover, concerning the collider signatures predictions, the radion evades current constraints but may show up in future LHC runs.
}

\FullConference{XIII Quark Confinement and the Hadron Spectrum - Confinement2018\\
		31 July - 6 August 2018\\
		Maynooth University, Ireland}

\begin{document}

\section{Introduction}
\label{sec:introduction}
\noindent The Standard Model (SM) of particle physics fails to explain a number of observational and theoretical issues, such as the baryon asymmetry of the universe, the origin of inflation and the strong sensitivity to high scale physics. The latter problem is commonly known as the hierarchy problem, and has motivated the study of several Beyond the SM (BSM) scenarios. One of the most fruitful BSM frameworks is the Randall-Sundrum model~\cite{Randall:1999ee}. In this scenario the hierarchy between the Planck and the ElectroWeak (EW) 
scale is generated by a warped extra dimension in Anti de Sitter (AdS) space. This model contains a light state, the radion, which is dual to the dilaton, a Goldstone boson of the conformal invariance of the dual 
four-dimensional (4D) theory, and it is typically the lightest BSM state. The radion undergoes a 'holographic' first order phase transition during which it acquires a vacuum expectation value~\cite{Creminelli:2001th,Nardini:2007me}. Models with small back-reaction on the gravitational metric suffer from perturbativity problems of the 
five-dimensional (5D) gravitational theory. In the present work we will introduce a method to deal with large back-reaction issues, that generalizes the superpotential procedure. 
For concreteness we will focus on  
a class of theories where the conformal symmetry is strongly broken in the infrared (IR). This kind of models were also considered to study EW precision data~\cite{Cabrer:2009we,Megias:2015ory}, and most recently the B-meson anomalies~\cite{Megias:2016bde,Megias:2017ove,Blanke:2018sro}.

\section{The radion effective potential}
\label{sec:radion_eff_pot}
\noindent 
In this section we 
present the model, and develop a novel method to employ the superpotential formalism to compute the effective potential at zero and finite temperature without major problems with the back-reaction. We consider a scalar-gravity system with two branes at values $r=r_0$ (UV brane), and $r = r_1$ (IR brane). The 
5D action of the model reads
\begin{eqnarray}
S &=& \int d^5x \sqrt{|\det g_{MN}|} \left[ -\frac{1}{2\kappa^2} R + \frac{1}{2} g^{MN}(\partial_M \phi)(\partial_N \phi) - V(\phi) \right]  \nonumber\\
&&\qquad - \sum_{\alpha} \int_{B_\alpha} d^4x \sqrt{|\det \bar g_{\mu\nu}|} \Lambda_\alpha(\phi)  
 -\frac{1}{\kappa^2} \sum_{\alpha} \int_{B_\alpha} d^4x \sqrt{|\det \bar g_{\mu\nu}|} K_\alpha \,.  \label{eq:S_model}
\end{eqnarray}
There are three kind of contributions to the action corresponding to
the bulk, the brane and the Gibbons-Hawking-York contribution.
$V(\phi)$ is the bulk potential, $\Lambda_\alpha(\phi) \;
(\alpha=0,1)$ are the UV and IR 
4D  
brane potentials at
$(\phi(r_0),\phi(r_1))$, and $\kappa^2 = 1/(2 M^3)$ with $M$ being the
5D Planck scale. The metric is defined in proper coordinates
as~$ds^2 = \bar{g}_{\mu\nu} dx^\mu dx^\nu-dr^2$ with $\bar{g}_{\mu\nu}
= e^{-2A(r)} \eta_{\mu\nu}$ the 4D induced metric on the branes. In
order to solve the hierarchy problem, the brane dynamics should fix
$(\phi_0,\phi_1)$ to get $A(\phi_1) - A(\phi_0) \approx 35$, and this
implies $M_{\Planck} \simeq 10^{15} M_{\textrm{TeV}}$. We have introduced a superpotential, whose relation with the scalar potential is
$V(\phi) = \frac{1}{8} W^\prime(\phi)^2 - \frac{\kappa^2}{6}
W^2(\phi)$. 

\subsection{The effective potential}
\label{subsec:radion_potential}
\begin{figure*}[t]
\begin{tabular}{cc}
\includegraphics[width=70mm]{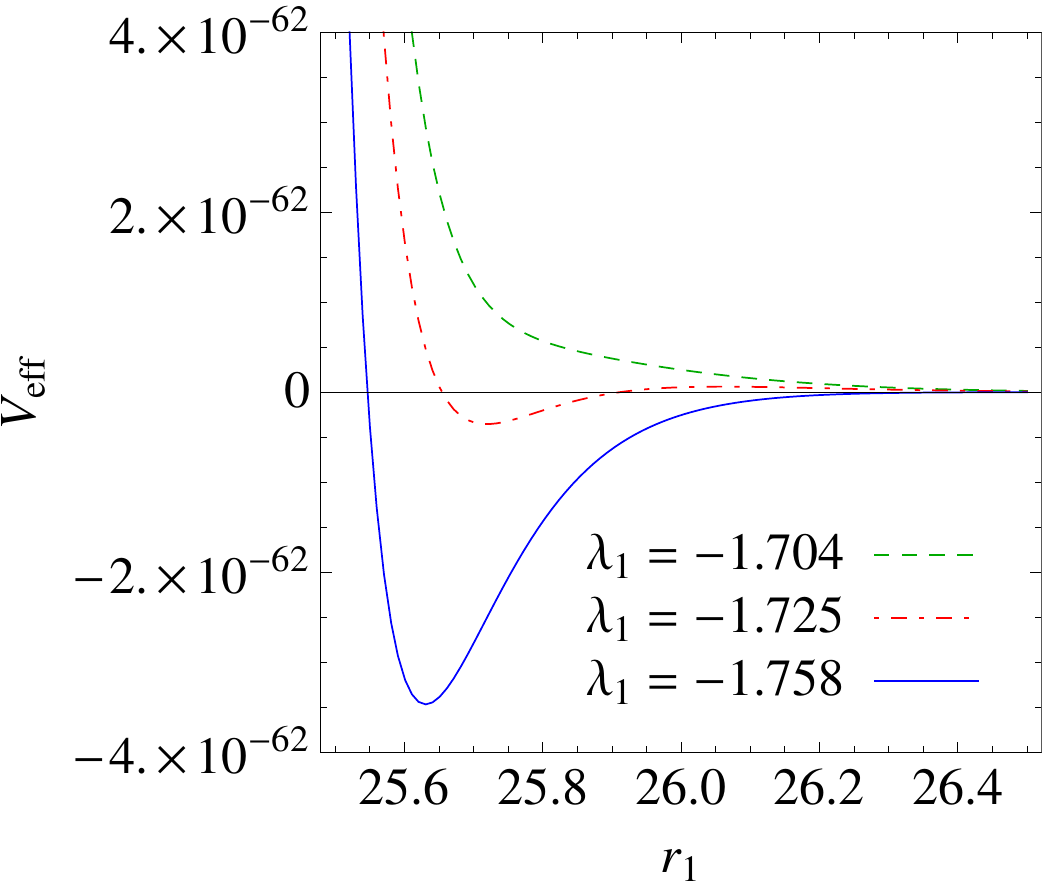} & \hspace{1cm}
\includegraphics[width=60mm]{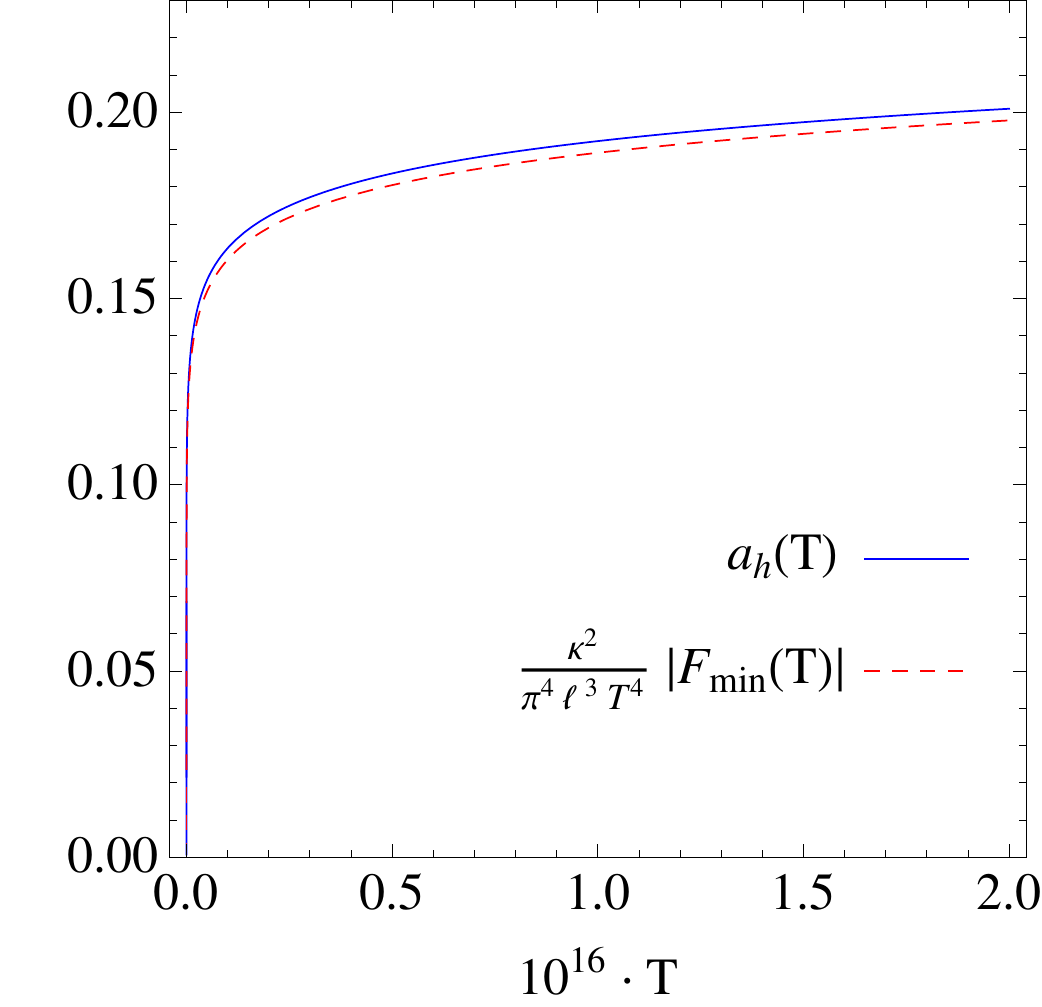} \\
\end{tabular}
\caption{Left panel: Effective potential for different values of $\lambda_1$ in units of $\ell$. Right panel: $a_h(T)$ (blue solid line) and $\kappa^2 |F_{\rm min}|/(\pi^4\ell^3T^4)$ (red dashed line) as a function of $T$. We rescale the brane tensions as~$\lambda_\alpha = \frac{\ell \kappa^2}{6} \Lambda_\alpha$. 
In both panels we 
have considered 
the 'large back-reaction scenario B'. }
\label{fig:Veff}
\end{figure*}

\noindent 
By using the equations of motion of the model, the action~(\ref{eq:S_model}) can be written as~\cite{Megias:2018sxv}
\begin{equation}
S  = - \int d^4x \,V_\eff   \,, \qquad \textrm{where} \qquad V_\eff =  \left[ e^{-4A} \left( W + \Lambda_1 \right) \right]_{r_1}   +  \left[ e^{-4A} \left( - W + \Lambda_0 \right) \right]_{r_0}  \,. \label{eq:Veff}
\end{equation}
After fixing $r_0 = 0$, the variable $r_1$ turns out to be the brane distance. The equation of the superpotential is a first order differential equation, so that it admits an integration constant that we will denote by~$s$. If $W_0$ is a particular solution of the equation of motion with potential~$V$, then it is possible to find a general solution of the equation as an expansion of the form $W=\sum_{n=0}^\infty s^n W_n$, where $W_n$ can be computed iteratively from $W_0$. An explicit solution for $n=1$ is given by~\cite{Papadimitriou:2007sj,Megias:2014iwa}
\begin{equation}
W_1(\phi)= \frac{1}{\ell\kappa^2} \exp\left(\frac{4\kappa^2}{3}\int_{v_0}^\phi \frac{W_0(\bar\phi)}{W'_0(\bar\phi)} d\bar\phi \right) \,,
\end{equation}
where $\ell$ is the AdS radius. We can similarly expand the scalar field as~$\phi(r) =\phi_0(r) + s \, \phi_1(r)+\mathcal O(s^2)$. The integration constant,~$s$, is fixed by the boundary condition $\phi(r_1) = v_1$, leading to~$s(r_1) = (v_1-\phi_0(r_1))/\phi_1(r_1)$. Therefore the superpotential 
acquires 
an explicit dependence on the brane distance, $W(v_\alpha) = W_0(v_\alpha) + s(r_1) W_1(v_\alpha) + \cdots$, which in turn creates a non-trivial dependence on $r_1$ of the effective potential of Eq.~(\ref{eq:Veff}). 
We will use this formalism for a kind of soft-wall phenomenological models defined by the superpotential~$W_0(\phi) = \frac{6}{\ell \kappa^2} \left(1 +  e^{\gamma \phi} \right)$ and concentrate in several benchmark scenarios, covering parameter configurations with large and small back-reactions on the metric. In particular: i) Scenario A (small back-reaction), $\kappa^2=\frac{1}{4}\ell^3$; ii) Scenario B (large back-reaction), $\kappa^2=\frac{1}{4}\ell^3$; and iii) Scenario C (large back-reaction), $\kappa^2=\ell^3$ (cf.~Ref.~\cite{Megias:2018sxv} for details). 
Table~\ref{tab:table} shows
the radion 
mass obtained 
in each scenario by using the approximate mass formula 
of Ref.~\cite{Megias:2015ory}.  Using this technique we find, for scenario B, the effective potential of Fig.~\ref{fig:Veff} (left). 
\begin{table}[htb!]
\centering
\begin{tabular}{||c|c|c|c|c|c||c|c|}
\hline\hline
Scen. & $\lambda_1$ & $m_\rad/\GeV$ & $T_c/\GeV$ & $T_n/\GeV$ & $T_R/\GeV$ &$\alpha$& $\log_{10}(\beta/H_\star)$ \\ \hline
A &  -1.250   & 199.8  & 230 & - & - & - & -  \\  \hline
B$_1$  &\textcolor{red}{-2.583} & 915 & 633 & 328  & 821.8 & 4.61  & 1.99 \\
B$_2$  &\textcolor{blue}{-2.125} & 745 & 463 & 26 & 566.4 & $4.0\cdot 10^4$  & 1.23  \\ \hline
C  & \textcolor{red}{-3.462} & 477  & 252  & 112 & 133.7 & 5.0  & 1.05  \\ 
\hline\hline
\end{tabular}
\caption{\it List of benchmark scenarios considered in this work. The outputs obtained in each scenario are presented from the third column on. The foreground \textcolor{red}{red} [\textcolor{blue}{blue}] color on the value of $\lambda_1$ indicates that the corresponding phase transition is driven by $O(3)$ [$O(4)$] symmetric bounce solutions, cf.~Sec.~\ref{subsec:dilaton_PT}.}
\label{tab:table}
\end{table}

\subsection{The effective potential at finite temperature}
\noindent
In the AdS/CFT correspondence a black hole (BH) solution describes the high temperature phase where the dilaton is in the symmetric phase, i.e.~$\langle \mu \rangle = 0$.~\footnote{We denote by $\mu(r)$ the canonically normalized radion field.} Let us consider the BH metric
\begin{equation}
ds_{\BH}^2 = -h(r)^{-1}dr^2 + e^{-2A(r)} (h(r) dt^2 -  d\vec{x}^{\,2} ) \,,  \label{eq:metricBH}  
\end{equation}
where the blackening factor vanishes at the event horizon, i.e.~$h(r_h) = 0$. The equations of motion of the system reduce to three independent equations with five integration constants, which can be fixed by imposing boundary conditions at the UV brane, $r=0$, and regularity conditions at the horizon, $r=r_h$. Then, the temperature and entropy of the BH can be expressed as
\begin{equation}
T_h = \frac{1}{4\pi} e^{-A(r_h)} \big| h^\prime(r) \big|_{r=r_h}  \,, \qquad S = \frac{4\pi}{\kappa^2} e^{-3A(r_h)} \equiv \frac{4\pi^4 \ell^3}{\kappa^2} a_h(T_h) T_h^3 \,, \label{eq:Ths}
\end{equation}
where~$a_h(T)$ is a smooth function which measures the deviation from the conformal limit, being $a_h = 1$ the conformal solution. Finally, the free energy of the system has a minimum at $T_h = T$, which can be approximated by~$F_{\mmin}(T) \simeq -\frac{\pi^4\ell^3}{\kappa^2} a_h(T) T^4$.  The numerical result of this procedure within scenario B is shown in Fig.~\ref{fig:Veff} (right). For small back-reaction $a_h(T)$ basically reproduces the case of pure AdS ($a_h = 1$), whereas for large back-reaction the value is $a_h(T) \ll 1$. This effect strongly influences the nucleation temperature of the phase transition. We find from our numerics that the behavior of $a_h(T)$ is generic and only depends on the amount of back-reaction.

\section{The phase transition}
\label{sec:phase_transition}
\noindent In this section we 
study 
the phase transition mechanism of the radion as well as its implications for the EW phase transition.

\subsection{The dilaton phase transition}
\label{subsec:dilaton_PT}
\begin{figure*}[t]
\begin{tabular}{cc}
\includegraphics[width=70mm]{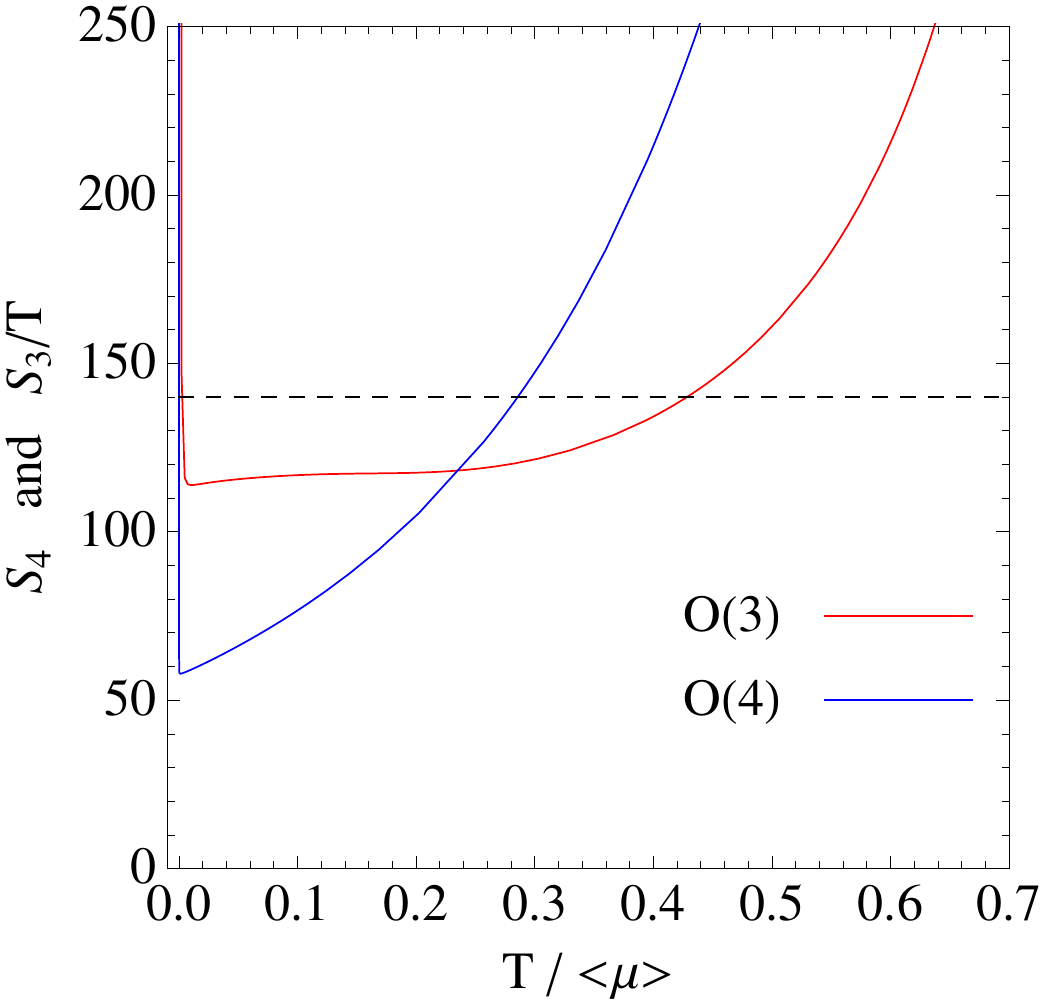} & \hspace{0.5cm}
\includegraphics[width=66mm]{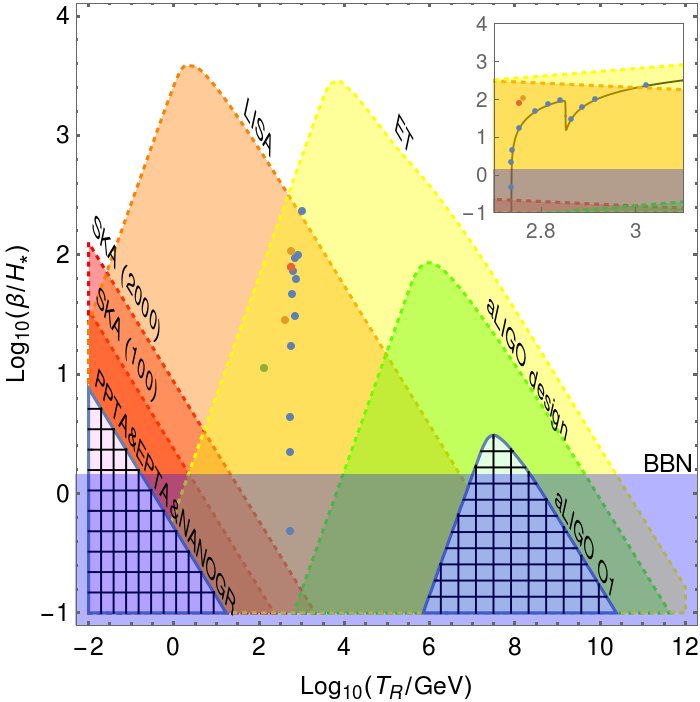} \\ 
\end{tabular}
\caption{Left panel: $S_4$ and $S_3/T$ as a function of temperature (normalized to $\langle \mu \rangle$). We have considered scenario B$_1$. Right panel: The $T_R - \beta/H_\star$ parameter space that exhibits  SNR $>10$ at SKA, LISA, aLIGO, and ET~\cite{TheLIGOScientific:2016dpb, Lentati:2015qwp, Arzoumanian:2018saf, Shannon:2015ect}. The Big Bang nucleosynthesis bound excludes the blue area. It is displayed as dots the values obtained in several benchmark scenarios: B (blue points), C (green point) and other scenarios discussed in Ref.~\cite{Megias:2018sxv}. The inserted figure is a zoom of the main one in the regime of the model prediction.}
\label{fig:ST}
\end{figure*}

\noindent 
When considering the system at finite temperature, there are two competing phases: the BH deconfined phase and the soft-wall confined phase, characterized by $\langle \mu \rangle = 0$ and $\langle \mu \rangle \ne 0$, respectively. The free energies of these phases are 
\begin{equation}
F_{\rm{deconfined}}(T) = E_0 + F_{\mmin}(T) -\frac{\pi^2}{90}g_d^{\eff} T^4 \,, \qquad F_{\rm{confined}}(T)=-\frac{\pi^2}{90}g_c^{\eff} T^4 \,,
\end{equation}
where $E_0 \equiv V_{\eff}(\mu=0)-V_{\eff}(\mu=\langle \mu \rangle)>0$, while $g_{c(d)}^{\eff}$ are the effective degrees of freedom in the confined (deconfined) phase. When the temperature is decreasing, the critical temperature at which the phase transition starts being allowed is given by $F_{\rm{deconfined}}(T_c) = F_{\rm{confined}}(T_c)$. The phase transition happens when the barrier between the false BH minimum and the true vacuum is overcome. While at high $T$ this process is driven by thermal fluctuations and the corresponding Euclidean action is $O(3)$ symmetric, at low~$T$ the transition can occur via quantum fluctuations 
with an $O(4)$-symmetric action. 
The corresponding equation of motion of the radion field is known as the 
`bounce equation' and, for the $O(n)$ symmetric Euclidean action, it is written as~\cite{Coleman:1977py}~\footnote{In these expressions $\rho = \sqrt{\vec{x}^2}$ for $n=3$, and  $\rho = \sqrt{\vec{x}^2 + \tau^2}$ (with $\tau$ being the Euclidean time) for $n=4$.}
\begin{equation}
\frac{\partial^2 \mu}{\partial\rho^2}  + \frac{(n-1)}{\rho} \frac{\partial \mu}{\partial\rho} - \frac{\partial V_{\eff}}{\partial\mu} = 0 \,, \qquad \textrm{with} \qquad   \frac{1}{2}\mu^{\prime\, 2}(\rho) \big|_{\mu=0}=\left|F_{\mmin}(T)\right| \,. \label{eq:bouncesol3}
\end{equation}
The bubble nucleation rate from the false BH minimum to the true vacuum
per Hubble volume $\mathcal V$ 
is~$\Gamma/\mathcal V \simeq e^{-S_E} \simeq  e^{-S_3/T} + e^{-S_4}$, so that it is dominated by the least action. Nucleation happens when the probability for a single bubble to be nucleated within one horizon volume is ${\cal O}(1)$, which 
corresponds to $S_E\lesssim 4 \log\left(M_{\Planck}/\langle\mu\rangle\right)\approx 140$~\cite{Konstandin:2010cd}. Fig.~\ref{fig:ST} (left) shows the numerical result in a scenario with large back-reaction (B$_1$). For large (small) values of $|\lambda_1|$ the phase transition is dominated by the $O(3)  (O(4))$ bounce. When 
the 
back-reaction is small neither $S_4$ nor $S_3/T$ reach the upper bound.
On the other hand, inflation starts when $E_0$ dominates the value of the energy density, i.e.~$\rho_{\rm{deconfined}}(T_i) \simeq E_0$, and it finishes when bubbles percolate at~$T \simeq T_n$. Finally, one can assume that during the phase transition the energy density is approximately conserved, so that at the end of the transition the universe ends up in the confined phase at the reheating temperature $T_R$ given by~$\rho_{\rm{confined}}(T_{R})=\rho_{\rm{deconfined}}(T_n)$. 
%
In Table~\ref{tab:table} we summarize the values of these observables.

\subsection{The electroweak phase transition}
\label{subsec:EW_PT}
\noindent
When $T$ decreases and the BH moves beyond the IR brane, the Higgs field 
$H$, localized nearby the IR brane, 
appears. Then the effective potential becomes a function of the radion and the Higgs~\cite{Nardini:2007me}:
%
\begin{equation}
V(\mu, H)=V_{\eff}(\mu)+\left(\frac{\mu}{\langle\mu\rangle}\right)^4V_{\SM}(H,T) \,, \quad \textrm{with} \quad V_{\SM}(H,T)=-\frac{1}{2}m^2 H^2+\frac{\lambda}{4} H^4 + \Delta V_{\SM}(H,T) \,.
\label{eq:totalpotential}
\end{equation}
Here  
the Higgs mass is $m_H^2=2\lambda v^2\simeq(125\,\GeV)^2$ with $v=246 \,\GeV$, and $\Delta V_{\rm SM}(H,T)$ contains loop corrections. $V_{\rm SM}(H,T)$ has its absolute minimum at $\langle H(T)\rangle = v(T)$ which, at leading approximation for the thermal corrections, turns out to be
\begin{equation}
v(T)=\left\{
\begin{tabular}{cl}
0 & for $T>T_{\EW}$  \\
$v\sqrt{1-T^2/T^2_{\EW}}$ & for $T\leq T_{\EW}$ 
\end{tabular}
\right. 
\end{equation}
with $T_{\EW} \simeq 150 \, \GeV$. The relevant quantity to study, when the universe ends up in the EW broken phase after the bubble percolation, is the reheating temperature. In scenarios with $T_R > T_{\EW}$, at the end of the reheating process the Higgs field is in its symmetric phase, and the EW symmetry breaking would occur as in the SM, i.e.~via a crossover that prevents the phenomenon of EW baryogenesis. Only in situations with $T_R < T_\EW$ the reheating does not restore the EW symmetry, and the Higgs lies at 
the minimum
$v(T_R)$. In the presence of a SM-like low energy particle content, the condition for EW baryogenesis~$v(T_R)/T_R \gtrsim 1$ is fulfilled when $T_R$ satisfies the bound~$T_R\lesssim T_H \simeq 140 \, \GeV$~\cite{Nardini:2007me, DOnofrio:2014rug}, otherwise the EW phase transition is too weak. A parameter configuration leading to~$T_R < T_H < T_\EW$ is provided by scenario~C, in which the EW and dilaton phase transitions happen simultaneously at $T = T_n = 112\, \GeV$, and end up with $T = T_R = 133.7\, \GeV$, cf.~Table~\ref{tab:table}.

\section{Gravitational waves}
\label{sec:GW}
\noindent
During a cosmological first order phase transition, a stochastic gravitational wave (GW) background is generated. The corresponding GW spectrum depends on several quantities that characterize the phase transition.
Within the envelope approximation, the frequency power spectrum of the stochastic GW background is given by~
\cite{Caprini:2015zlo}
\begin{equation}
h^2\Omega_{\rm GW}(f) \simeq  h^2\overline \Omega_{\rm GW} ~\frac{3.8(f/ f_p)^{2.8}}{1+2.8 (f/ f_p)^{3.8}}  \,, \qquad \textrm{with} \qquad  h^2\overline \Omega_{\rm GW} \propto \left( \frac{H_\star}{\beta}   \frac{\alpha}{\alpha+1} \right)^2 \,,
\label{eq:OmGW}
\end{equation}
where $H_\star$ is the Hubble parameter at the time $t_\star$ when
the bulk of the GW production starts.  
There are two parameters, $\alpha$ and $\beta$, that
control the behavior of the spectrum. The parameter $\alpha$ is
related to the latent heat of the phase transition, while $\beta
\equiv - dS/dt|_{t=t_\star}$ is a measurement of the time duration of
the phase transition. These two quantities can be computed as
\begin{equation}
\alpha \simeq\frac{E_0}{3(\pi^4\ell^3/\kappa^2) a_h(T_n) T_n^4} \,, \qquad \textrm{and} \qquad  \frac{\beta}{H_\star}\simeq  T_n\left. \frac{dS_E}{dT}\right|_{T=T_n} \,, \label{eq:alphabetaGW}
\end{equation}
where it has been assumed that $T_\star \approx T_n$. A relevant quantity that measures the capability of an experiment to detect GWs is the Signal-to-Noise Ratio (SNR). It is proportional to $\SNR \propto \sqrt{\mathcal{T}}$, where $\mathcal{T}$ is the time period collecting data, and it depends on the sensitivity curve of the experiment in frequency space. Fig.~\ref{fig:ST} (right) shows the parameter region in the plane $T_R - \beta/H_\star$ that exhibits $\SNR > 10$ for different experiments over the next $\sim 20$ years. Note that in a wide fraction of this parameter space, at least two experiments will be able to detect independently the signals~\cite{Figueroa:2018xtu}. The results obtained with the benchmark scenarios of the model presented in Secs.~\ref{sec:radion_eff_pot} and \ref{sec:phase_transition} are displayed as dots, and they turn out to be detectable at both LISA and Einstein Telescope (ET) experiments.

\section{Heavy radion phenomenology}
\label{sec:heavy_radion_phenomenology}
\noindent
Finally, we 
study the detection prospects for the radion in collider phenomenology at the LHC. In our particle setup, when the radion is lighter than any KK resonance and $m_{\textrm{rad}} \sim {\cal O}(\textrm{TeV})$, it can decay only into SM-like fields. Since the radion couples to the 
energy-momentum tensor, its production/decay channels are those of the SM Higgs, but with different strengths. We assume that the Higgs is exactly localized at the IR brane. Using the 4D action for the radion and SM fields, one can compute the couplings of the radion to the massless gauge bosons (photons and gluons), massive gauge bosons ($W^\pm$ and $Z^0$), fermions and the Higgs boson. As an example, the coupling to fermions looks like~$\mathcal L_{r\bar ff}=-\frac{\mathcal R(x)}{v}c_f m_f \bar f f$, where $c_f$ is a coefficient which measures the departure of the ${\bar f}f$ coupling from its value for a hypothetical SM Higgs $H$ with $m_H = m_\rad$, cf.~Ref.~\cite{Megias:2018sxv} for full details. We show in Table~\ref{tab:tablecouplings} the numerical values of the coupling coefficients.
\begin{table}[t]
\centering
\begin{tabular}{||c||c|c|c|c|c|c|c||}
\hline\hline
        & $c_\gamma$&$c_g$&$c_V$ &$c_H$& $c_f$ & & \\ \hline
        & 0.472  & 0.164 & 0.0649  & 0.259  & 0.259 & &  \\ \hline\hline
        & $B^{\mathcal R}_{WW}$ & $B^{\mathcal R}_{ZZ}$ & $B^{\mathcal R}_{HH}$& $B^{\mathcal R}_{t\bar t}$& $B^{\mathcal R}_{b\bar b}$& $B^{\mathcal R}_{\tau\bar \tau}$& $B^{\mathcal R}_{\gamma\gamma}$\\ \hline

      & 0.271  &  0.135   & $1.26 \cdot 10^{-3}$   & 0.592  & $1.83 \cdot 10^{-4}$  &  $2.85 \cdot 10^{-5}$  &  $8.55 \cdot 10^{-6}$   \\  \hline\hline
     & $S^{\ggF}_{WW}$ & $S^{\ggF}_{ZZ}$ &  $S^{\ggF}_{\tau\bar \tau} $ & $S^{\ggF}_{\gamma\gamma}$+$S^{\VBF}_{\gamma\gamma}$ &  $S^{\VBF}_{WW}$ & $S^{\VBF}_{ZZ}$ &  $S^{\VBF}_{\tau\bar \tau} $
\\ \hline

(predic.) & 1.59   &  0.80   & $1.7 \cdot 10^{-4}$   &  $5.5 \cdot 10^{-5}$  & 0.16  & 0.080  &$1.7 \cdot 10^{-5}$  \\ \hline
(bound) &  52 & 14   & 11   & 0.29  & 12  & 8  &     -- \\
\hline\hline
\end{tabular}
\caption{\it  Coupling coefficients of the radion interactions with the SM fields, radion branching fractions, and the predictions of $S^{ggF(VBF)}_{XX}$  and their corresponding 95\% C.L.~upper bounds (in fb units). The bounds are taken from the ATLAS searches, see e.g.~Ref.~\cite{Aaboud:2017fgj}. We have considered the scenario B$_1$, cf. Table~\ref{tab:table}.
}
\label{tab:tablecouplings}
\end{table}

The main production mechanisms of the heavy radion at the LHC are gluon fusion (ggF)~and vector-boson fusion (VBF). Noting that the production cross-sections~$\sigma_\mathcal{R}(gg(VV) \to \mathcal R) \propto |c_{g(V)}|^2$, and the decay widths of the radion into SM particles~$\Gamma(\mathcal R \to X\bar X) \propto |c_X|^2$, we can compute these quantities from their relations to the heavy SM Higgs predictions, thus leading to the cross sections
\begin{equation}
S_{XX}^{\ggF(\VBF)} \equiv \sigma^{\ggF (\VBF)}(pp \to \mathcal{R} \to X{\bar X}) \simeq  \sigma_\mathcal{R}^{\ggF (\VBF)}(gg(VV) \to \mathcal{R})  \cdot B^{\, \mathcal R}_{XX}(\mathcal{R} \to X{\bar X})   \,,
\end{equation}
where $B^{\, \mathcal R}_{XX}$ are the radion branching fractions with $X=\gamma,W,Z,f, H$. We summarize in Table~\ref{tab:tablecouplings} the results in the benchmark scenario B$_1$. We conclude that this scenario is in full agreement with the current bounds at the LHC. The same conclusion is obtained when studying other scenarios, cf.~Ref.~\cite{Megias:2018sxv}. In particular, scenario C has channels $ZZ$ and $WW$ that are not far below the experimental constraints, so that future LHC data will be able to probe some of these decay channels. This will probably happen in conjunction with the LISA and ET measurements.

\section{Conclusions}
\label{sec:conclusions}
\noindent 
In the context of 
5D warped models, we have introduced a novel technique to compute the effective potential of the radion based on a suitable perturbative expansion of the general solution of the 
equations of motion for the superpotential. This method can be applied to any warped model, even in situations with a strong back-reaction over the metric. From the computation of the effective potential at finite temperature using a black hole solution, we have been able to study the dilaton phase transition, and obtain that the confinement/deconfinement phase transition demands a sizable large back-reaction. We
have
 also 
identified 
some scenarios in which the EW symmetry breaking happens at the same time that the dilaton phase transition, and EW baryogenesis is still possible. The model predicts the production of gravitational waves in a regime detectable either by LISA and the Einstein Telescope. Finally, we have found that 
the current LHC bounds are not in tension with the (rather heavy) radion predicted by the model.


\end{document}